\documentclass[preprint2]{aastex631}
\usepackage{longtable}
\usepackage{amsmath}
\usepackage{todonotes}
\usepackage{subfigure}
\usepackage{multirow}

\shorttitle{Voronoi Tesselations and Radial Density Profiles}
\shortauthors{Dornan \& Harris}

\graphicspath{{./}{figures/}}

\begin{document}


\title{Utilizing Voronoi Tessellations to Determine Radial Density Profiles} 

\author[0000-0002-7731-1291]{Veronika Dornan}
\affiliation{Department of Physics and Astronomy \\
McMaster University \\
Hamilton, ON, L8S 4M1}

\author[0000-0001-8762-5772]{William E. Harris}
\affiliation{Department of Physics and Astronomy \\
McMaster University \\
Hamilton, ON, L8S 4M1}

\begin{abstract}

We have developed a novel method of determining 2D radial density profiles for astronomical systems of discrete objects using Voronoi tessellations. This Voronoi-based method was tested against the standard annulus-based method on 5 simulated systems of objects, following known Hubble density profiles of varying parameters and sizes. It was found that the Voronoi-based method returned radial density fits with lower uncertainties on the fitting parameters across all 5 systems compared to the annulus-based method. The Voronoi-based method also consistently returned more accurate estimates of the total number of objects in each system than the annulus-based method, and this accuracy increased with increasing system size. Finally, the Voronoi-based method was applied to two observed globular cluster systems around brightest cluster galaxies ESO 444-G046 and 2MASX J13272961-3123237 and the results were compared to previous results for these galaxies obtained with the annulus-based method. Again, it was found that the Voronoi-based method returned fits with lower uncertainties on the fitting parameters, and the total number of globular clusters returned are within errors of the annulus-based method estimates, however also with lower uncertainties.   

\end{abstract}


\section{Introduction} \label{sec:intro}

Determining the radial density profiles of a variety of astronomical systems that consist of discrete objects is a frequently encountered problem within the field. Prominent examples may include stars within a star cluster \citep{king1966, Miocchi2013}, systems of globular clusters (GCs) within their host galaxy \citep{Harris15, Dornan23}, or galaxies within a larger group or cluster \citep{nagai2005}. Thus, in order to get accurate estimates of the central concentration, profile shape, and total number of objects within a system it is necessary to have an accurate fit to the projected radial density profile.

Observationally, the classic method to determine these radial density profiles is to first determine the 2-dimensional spatial distribution of objects from photometric observations, then divide this spatial distribution into a series of concentric circular or elliptical annuli, effectively binning the objects as a function of radius. For each of these bins the number of objects and the area of the bin's annulus are determined in order to obtain a density value. These density values are then plotted as a function of radius to which a density profile can then be fit. This method had been applied observationally to stars in open star clusters \citep[e.g.][]{alfonso2023, open1, open2}, stars in GCs \citep[e.g.][]{Knodlseder2000, dalessandro2015, simpson2017}, systems of GCs around galaxies \citep[e.g.][]{Harris15, Harris2016, Mulholland17,Dornan23,caso2024,Ennis24}, galaxies within galaxy clusters \citep[e.g.][]{Popesso07,Newman13,Fassbender14,Tozuka21}, and compact radio sources at the centre of the Galaxy \citep{Zhao22}.

This annulus fitting method is useful in its simplicity, but has some drawbacks. The systems to which it is applied can have thousands of detected objects within this spatial distribution \citep{Peng2008, Peng2011, Harris2016, Dornan23}, which when binned, are reduced to only a few datapoints, resulting in a loss of information about the system, and uncertainty in the fit. In addition, the number of these binned annuli, and by extension the radial width they encompass, is arbitrary, which can introduce further uncertainty in the resulting fit. Finally, as well the annulus method prescribes a geometrically regular (circular or elliptical) density distribution, smoothing over substructure such as satellites or local overdensities.

This work introduces a different general methodology using Voronoi tessellations for determining radial density profiles. 
Voronoi tessellation plots are constructed from a spatial distribution of points, and each tessellation (Voronoi cell) consists of the area of the distribution that is closer to the given point than the other points \citep{Okabe92}. The result is a distribution of polygons each containing one of the objects from the input spatial distribution. See Figure \ref{fig:dists} for a simple example of a 2D Voronoi tessellation plot. 


The tessellation process allows for construction of the density distribution, since the area of a Voronoi cell is directly related to the local number density of objects surrounding the cell. Higher density regions will produce Voronoi cells with smaller areas, and lower density regions will produce Voronoi cells with larger areas. Thus, once a Voronoi tessellation plot has been created from a spatial distribution of objects, the area of each cell can be inverted to obtain a density value for each object. Dividing the region into Voronoi cells is a binning process like the standard annular method, but it essentially carries the definition of bin size to its minimum possible extreme, with one object per bin. In addition, unlike the classic annulus method, the Voronoi method makes no assumptions about any geometric regularity in the spatial distribution. 

It should be noted however, that in order to determine surface density using both annuli and Voronoi tessellations, a distribution of discrete objects is required. Neither of these methods can be applied to, for example, surface brightness profiles.

Voronoi tessellations have been applied to the determination of spatial densities in astronomical research in the past, as they can be very useful in describing the spatial distribution of observed galaxies, which can exist in areas of over and under densities \citep{Icke89, Icke91}. Voronoi tessellations have also been used in algorithms to automatically detect these over-densities, galaxy clusters, from 2D galaxy field spatial distributions \citep{ramella2001}. 

In the present paper, we test this Voronoi-cell method against simulated radial distributions of populations following a simple input distribution (a Hubble model profile), and compare the results with the standard annular-bin method. The data used to test this new Voronoi method will be discussed in section \ref{sec:data}. This section will describe the simulated 2D Hubble distributions and test the method's ability to return the input radial density profile, compared to the annulus method. Section \ref{sec:methods} will further describe the Voronoi method in detail and the statistical tests used to compare it to the annulus method. Section \ref{sec:results} will summarize the results of these statistical tests and outline the benefits of the Voronoi method for determining radial density profiles. Finally, section \ref{sec:discussion} will provide a sample application
to observed systems and discuss future applications of this method and improvements that can be made.


\section{Data} \label{sec:data}

In order to test the ability of both the annulus fitting method and the Voronoi fitting method to accurately return the true radial density profile of a 2D spatial distribution of objects, a series of artificial systems were created with known Hubble profiles, defined by

\begin{equation}\label{eq:hubble}
    \sigma = \sigma_0 \Big(1 + \frac{r}{r_0}\Big)^{-b} \, 
\end{equation}
where $\sigma_0$ is the density at $r=0$, $r_0$ is the core radius, and the exponent $b$ sets the power-law slope of the profile at large radius.  

We generated simulated systems with five different total populations, The properties of which are detailed in Table \ref{tab:art_hub} (The scale units are arbitrary, but are simply labelled as `pixels' for convenience). The spatial distribution of objects for each of these systems was generated with 30 different random seeds each, allowing for each system to be tested with the same underlying Hubble distribution but varying positions of objects.

The parameters for the artificial systems were chosen building from our previous work on the estimated properties of a sample of observed GC systems hosted by massive elliptical galaxies. These galaxies have already had the radial density profiles of their GC systems determined with the annulus fitting method in \cite{Dornan23}, hereafter referred to as Paper I. The parameters of the simulated systems were chosen to reflect the range of profiles seen in a real-life application of this methodology. 

It should be noted that in this initial study the Voronoi method is tested only on  circularly symmetric systems, although does not make any assumptions on the underlying symmetry of the system. This decision was made because the annulus method, to which we are directly comparing the Voronoi method, does. 
In followup work, the Voronoi method will be extended to systems where one or more satellites are present, with their own subpopulations of objects, 
in order to investigate the technique's ability to handle asymmetrical systems.


\begin{center}
\begin{table*}[h!tb]
    \caption{Artificial System Hubble Profile Parameters} \label{tab:art_hub}
    \begin{tabular}{ccccccc}
    \hline \hline
    System Number & Objects & Core (pixels) & $b$ & $r_{max}$ (pixels) & $\sigma_o$ (pixels$^{-2}$)\\
    (1) & (2) & (3) & (4) & (5) & (6) \\
    \hline
    1 & 300 & 75 & 1.5 & 3250 &  $8.83 \times 10^{-4}$\\
    2 & 2000 & 150 & 0.96 & 3250 & $6.93 \times 10^{-4}$ \\
    3 & 4500 & 150 & 0.96 & 3250 & $1.56 \times 10^{-3}$ \\
    4 & 7000 & 250 & 1.09 & 3250 & $2.05 \times 10^{-3}$ \\
    5 & 20000 & 250 & 1.09 & 3250 & $5.86 \times 10^{-3}$ \\
    \hline
    \end{tabular}
\item{} \footnotesize{\textit{Key to columns:} (1) Artificial system identification; (2) number of objects in the system; (3) size of the core of the Hubble profile in units of pixels; (4) exponent of the Hubble profile; (5) maximum radius objects can be from the centre of the spatial distribution in units of pixels; (6) density at $r=0$ of the Hubble profile in units of objects per pixel$^2$.}
\end{table*}
\end{center}

\section{Methods} \label{sec:methods}

\begin{figure*}[h!tb]
    \begin{center}
    \includegraphics[width=0.49\textwidth]{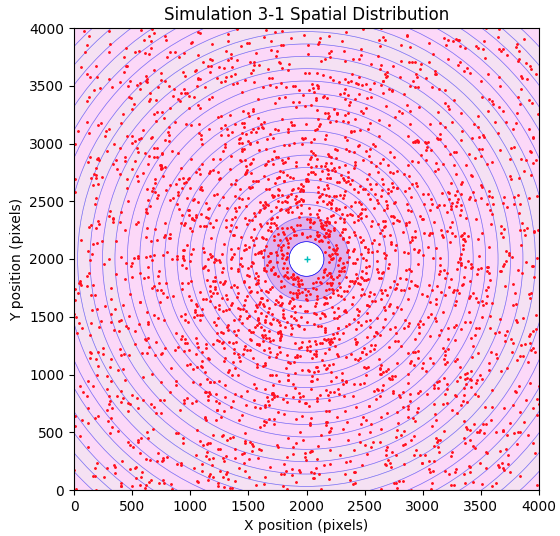}
    \includegraphics[width=0.49\textwidth]{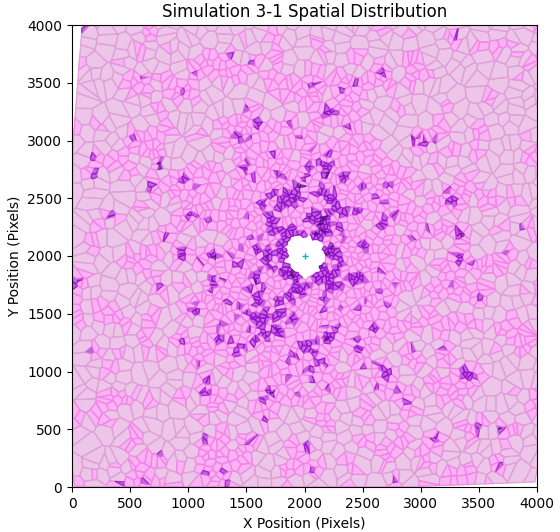}
    \caption{\label{fig:dists} \textit{Left:} The 2D spatial distribution of objects with concentric annuli drawn out from the centre of the system. These annuli are used to bin the objects and determine densities as a function of radius. \textit{Right:} The Voronoi Tessellation plot for the same 2D distribution of objects, with one object per cell. For both plots cell colour represents comparative density on the same scale.}
    \end{center}
\end{figure*}

\subsection{Annulus Fitting Method} \label{subsec:annulus}

As noted above, annulus fitting is a standard binning method used to determine radial density profiles of a variety of circularly and spherically distributed systems, both simulated and observed \citep{king1966, nagai2005, Miocchi2013, Dornan23, alfonso2023}.

An example of a 2D spatial distribution of objects divided into concentric annuli can be seen in Figure \ref{fig:dists}, which is for simulated system number 3. These binned densities can then be plotted as a function of radius and a simple fitting function such as a power law or Sersic profile can be fit to it.

For these simulated systems the number of annuli was always set to 26 and the bin widths determined from the range of radii encompassing the innermost and outermost objects. However, often the 1-2 outermost bins had to be removed for incompleteness, as often the final few annuli would only host a very low number of objects and areas. This can be seen in Figure \ref{fig:dists} in the corners of the annulus distribution.

\subsection{Voronoi Fitting Method} \label{subsec:voronoi}

The Voronoi fitting method is capable of retaining far more information  about the system than the annulus fitting method through the use of the Voronoi tessellations. When a Voronoi cell is generated
around each object in the system, the number of datapoints is equal to the number of objects within the system. A density value for each datapoint, at the radius of each object, is determined by simply inverting the area of the Voronoi cell.

\begin{figure}
    \begin{center}
    \includegraphics[width=0.46\textwidth]{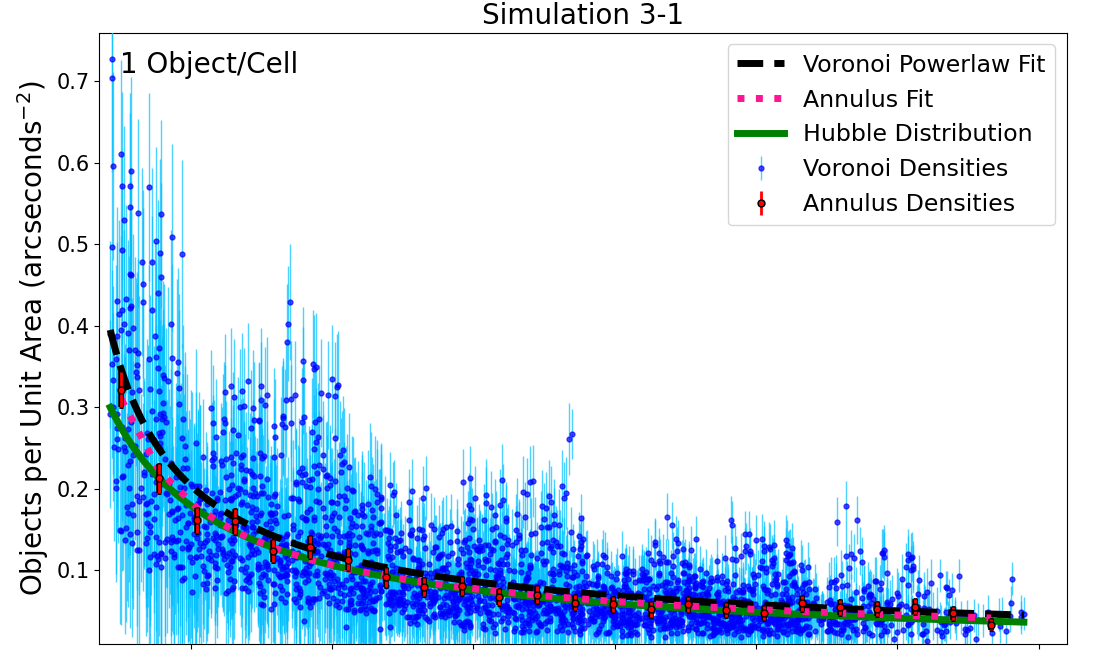}
    \includegraphics[width=0.46\textwidth]{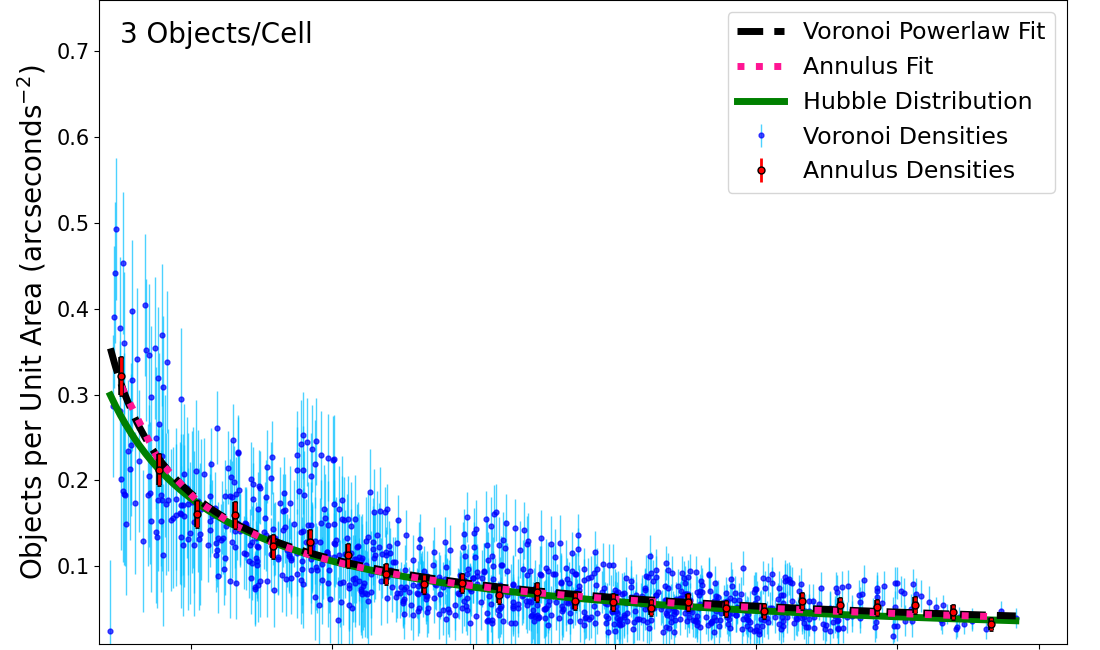}
    \includegraphics[width=0.46\textwidth]{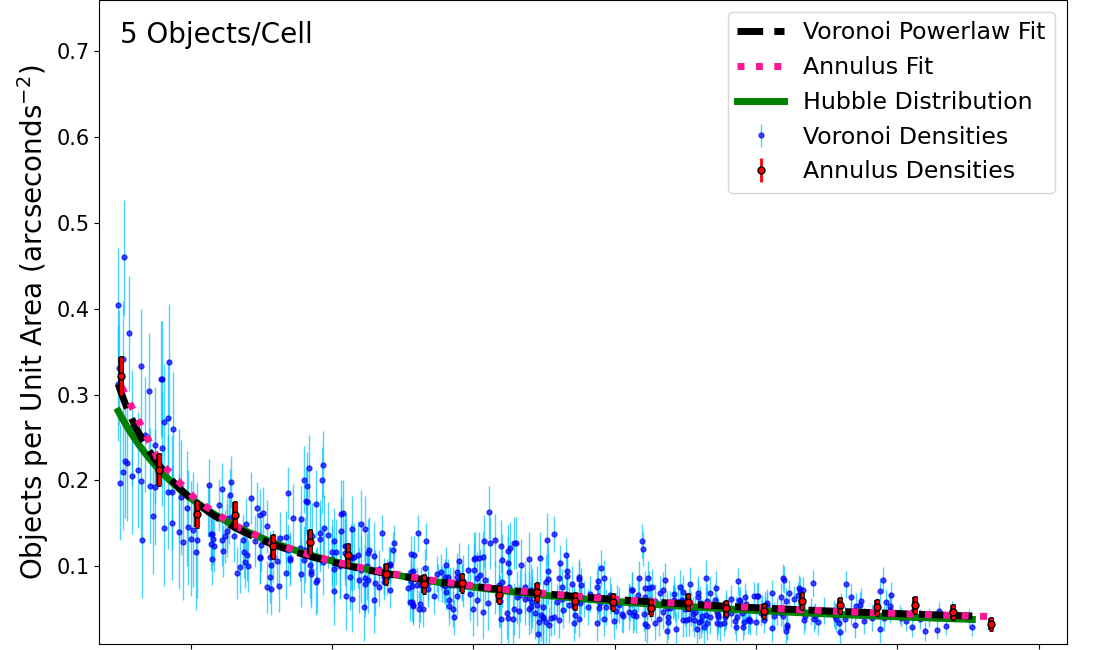}
    \includegraphics[width=0.46\textwidth]{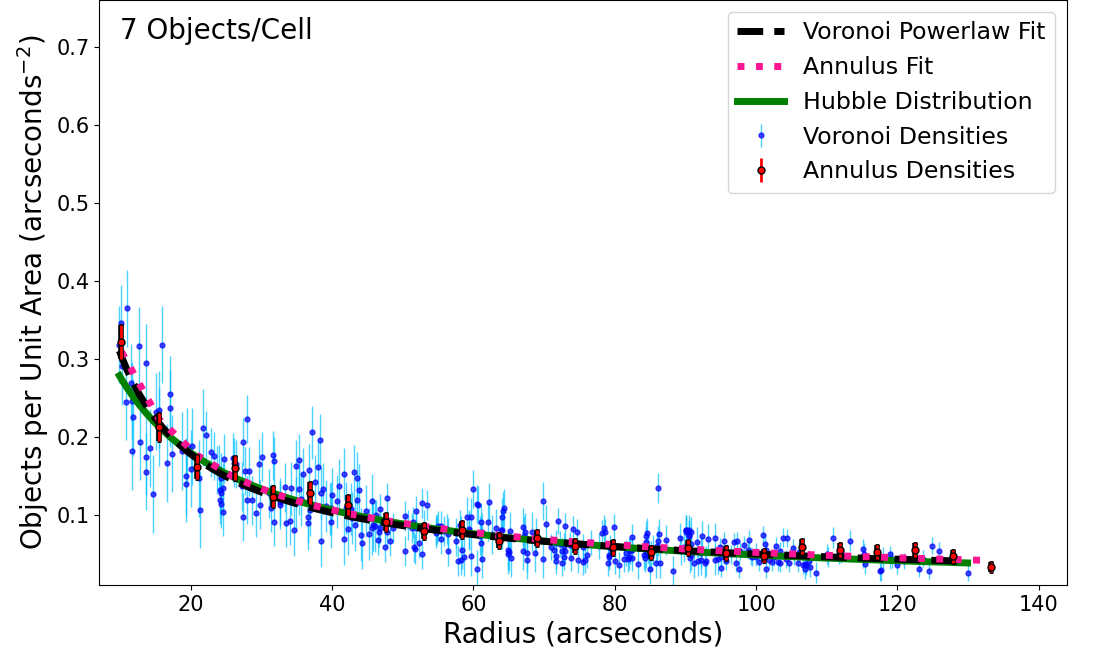}
    \caption{\label{fig:cell_nums} Sample radial profile showing the radial density distributions for 1, 3, 5, and 7 objects per cell (going from top down). All annulus data and fitting is identical, as it is independent from the Voronoi fitting.}
    \end{center}
\end{figure}

In practice, the resulting plot of density versus radius for one object per cell displays the desired profile, but is also dominated by large stochastic scatter.  A useful approach to reduce the scatter is simply to combine adjacent cells into larger combined ones. We have experimented with grouping 3, 5, or 7 cells together compared to no grouping at all, as shown in Figure \ref{fig:cell_nums}. Binning the Voronoi cells to have five objects per cell proved to be an excellent compromise between significantly reducing scatter and keeping a large number of independent datapoints. 

An alternative approach to grouping cells would be to combine a given cell with all its adjacent cells (i.e. those whose sides are in direct contact). 
However, for a typical Voronoi tesselation, approximately $89\%$ of all cells have 6 or more adjacent cells \citep[see Table 1 of][]{vor_sides}. Thus the majority of cells in the distribution would be binned in groups of 7 or more, which, as previously mentioned, was found to result in too strong an averaging and greater culling of datapoints.

Combining cells was done by sorting the cells by density, from highest to lowest. The highest density cell in the list then has the four spatially closest objects to it identified, and all five are added into a single bigger cell and removed from the initial list. This new cell is assigned an (x, y) position in the grid that corresponds to the averaged positions of the five member objects, and an area equal to the combined areas of the original cells. The rest of the binning continues in the same way for the next highest density individual cell in the list. This procedure ensures that the objects that are being binned together are spatially close to one another and prevents situations where the last two objects are on opposite sides of the system, or oddly shaped (``gerrymandered") combined cells are created.

\begin{figure}
    \begin{center}
    \includegraphics[width=0.45\textwidth]{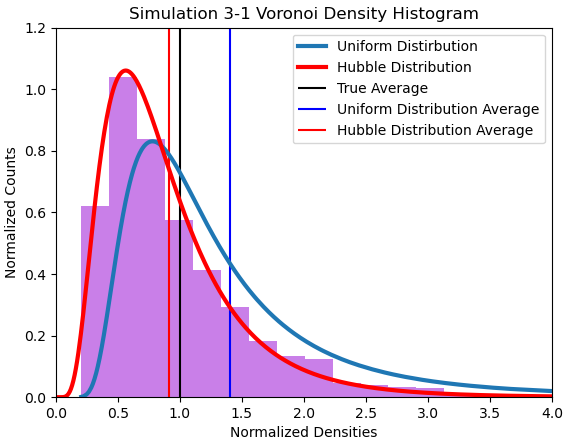}
    \caption{\label{fig:hist} A histogram of the normalized density of Voronoi cells for a Hubble distribution. The red curve is the log-normal fit to the histogram and the red line is the average for the Hubble distribution. The blue curve is the fit to the histogram for a uniform distribution and the blue line is average for the uniform distribution. The black line is the mean normalized number density, where $x=n/\langle n \rangle = 1$ when $n=\langle n \rangle$ for all distributions}.
    \end{center}
\end{figure}

Binning the cells together not only reduces scatter, but also reduces a particular statistical bias that is associated with Voronoi tessellations. A tessellation of a random, \emph{spatially  uniform} 2D distribution of objects can be described \citep{voronoi_bias} as

\begin{equation} \label{eq:bias_size}
    f_{2D}(y) = \frac{343}{15}\sqrt{\frac{7}{2\pi}}y^{5/2}\exp{\Big(-\frac{7y}{2}\Big)}
\end{equation}
where $y$ is the normalized cell area, $S$, as $y=S/\langle S \rangle$. This distribution for cell size can be transformed into a distribution for a number density ($n$) distribution for each cell. Here we will take $x = n/\langle n \rangle  = 1/y$ to be the normalized number density. The 2D number density distribution for a uniform spatial distribution of objects is 

\begin{equation} \label{eq:bias_density}
    f_{2D}(x) = \frac{343}{15}\sqrt{\frac{7}{2\pi}}x^{9/2}\exp{\Big(-\frac{7}{2x}\Big)}
\end{equation}

For this distribution the mean value of the probability density function does not equal the mean normalized number density, $x = n/\langle n \rangle$. It is instead 1.4 times the mean normalized number density, indicating that, for a uniform spatial distribution, Voronoi tessellations tend to overestimate the influence of high density areas for the overall system.  This can be seen in Figure \ref{fig:hist} when comparing the blue curve, which corresponds to Equation \ref{eq:bias_density}, and the blue vertical line, which represents the curve's mean, to the black line which is the expected mean value, where $n=\langle n \rangle$ and $x=1$.

However, for the purposes of this work we are not concerned with the number density bias for a uniform distribution, but rather for a distribution following a power-law-like radial profile. When the normalized number density distribution is plotted as a histogram for the artificial systems in this work, the distribution can instead be described by a log-normal distribution, taking the form:

\begin{equation} \label{eq:hub_hist}
    f_{2D}(x) = \frac{1}{x a \sqrt{2\pi}}\exp{\Big(\frac{(\log{x}-b)^2}{2 a^2}\Big)}
\end{equation}

Where $a$ and $b$ are free parameters. This can again be seen in Figure \ref{fig:hist}, where the red curve corresponds to Equation \ref{eq:hub_hist} and the red vertical line represents this curve's mean.

For a Hubble distribution, the average number density of the cells, after binning into groups of five, is actually 0.90 times the mean normalized number density. This statistical underestimation of the average number density of the Voronoi tessellations was found to be small enough that it has a negligible effect on the radial density profile fits. As will be seen below, the over- and under-estimations of the density profile at various radii due to simple randomness across the systems studied were consistently greater than the influence of this statistical bias, so at this time it is ignored.

\section{Results} \label{sec:results}

\begin{figure*}
    \begin{center}
    \includegraphics[width=0.95\textwidth]{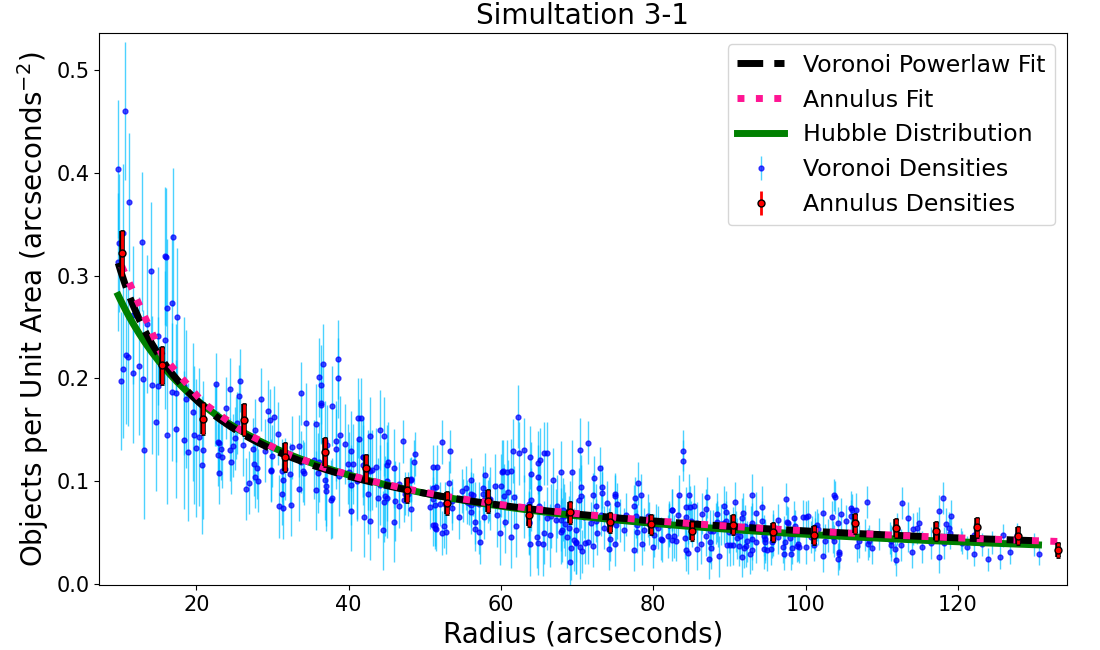}
    \caption{\label{fig:dense} The density data for the same system using the annulus fitting method (red datapoints) and the Voronoi fitting method (blue datapoints). Both datasets are fit with a powerlaw function and compared to the ``true" Hubble profile (green line).}
    \end{center}
\end{figure*}

\begin{center}
\begin{table*}[h!tb]
    \caption{Bootstrapped Powerlaw Fits with Standard Errors} \label{tab:stats}
    \begin{tabular}{|c|c|c|}
    \hline \hline
    System Number & Annulus Method & Voronoi Method \\
    \hline
    1 & $\sigma =  (1.131 \pm 0.110) r^{(-1.360 \pm 0.021)}$ & $\sigma =  (1.184 \pm 0.084) r^{(-1.370 \pm 0.020)}$\\
    2 & $\sigma =  (0.852 \pm 0.035) r^{(-0.794 \pm 0.010)}$ & $\sigma =  (0.849 \pm 0.029) r^{(-0.784 \pm 0.007)}$\\
    3 & $\sigma =  (1.899 \pm 0.040) r^{(-0.782 \pm 0.006)}$ & $\sigma =  (1.704 \pm 0.012) r^{(-0.753 \pm 0.002)}$\\
    4 & $\sigma =  (2.637 \pm 0.127) r^{(-0.757 \pm 0.010)}$ & $\sigma =  (2.765 \pm 0.011) r^{(-0.758 \pm 0.001)}$\\
    5 & $\sigma =  (7.377 \pm 0.244) r^{(-0.747 \pm 0.008)}$ & $\sigma =  (8.726 \pm 0.013) r^{(-0.780 \pm 0.0004)}$\\
    \hline
    \end{tabular}
\item{} \footnotesize{\textit{These fitting parameters and standard errors were obtained from 5000 iterations of bootstrapping for both methods.}}
\end{table*}
\end{center}

For each run of each system the density profiles were determined both with the annulus method and the Voronoi method. This created two different radial density datasets from the same underlying Hubble distribution. For both the annular and Voronoi methods, objects within the core of the system (defined in table \ref{tab:art_hub}) were removed. This is because in real observations of systems of globular clusters the background light intensity within the core of the central galaxy is typically high enough that identification and measurement of objects in the core cannot be done
\citep{Harris15, Harris2016, Harris23, Dornan23, Ennis24}. Thus, these objects have been removed for the analysis in order to better re-create the data available for a real astronomical system, to which this method will be applied in section \ref{subsec:real}.  

Figure \ref{fig:dists} shows the spatial distribution of the simulated objects and a visualization of the density fitting method used. Once the objects have been binned according to the methods used their densities can be plotted as a function of radius and have profiles fit to them. Here the radii and densities have been converted from pixels to ``arcseconds" according to HST ACS camera pixel conversions, defined as 1 pixel width $= 0.05$ arcseconds \citep{ACS}, again to better reflect observed data. The uncertainties associated with the annulus fitting method density values assume $\sqrt{N}$ statistics. 

The uncertainties of the density values from the Voronoi fitting method were determined from the root mean square scatter around a powerlaw fit to the densities as a function of radius, for 10 radial bins. This allows the decreasing uncertainty in cell density to be reflected in the voronoi cells, as is seen for the annulus bins. Once the uncertainty for each radial bin was determined it was applied to all density values in that bin.


Once the uncertainties for both methods are determined the Voronoi fitting method density values were culled to remove outliers. This is done by determining the average Voronoi density value for the same number of radial bins as was used for the annuli (although the average Voronoi densities in these bins are slightly different than the annuli density values). All Voronoi density values that are within 1.5 standard deviations of these radially binned means are kept, and all others are culled. The percentage of cells kept after culling varied for system and run, but was roughly 85\%.

In the case of this work, although the distribution was created according to a Hubble profile, the densities are fit with a simpler powerlaw function due to the lack of information within the core. These fits can then be compared to the ``true" Hubble profile to determine which method returns the more nearly correct density profile. An example is shown in Figure \ref{fig:dense}. If this Voronoi method was applied to, for example, a single GC, a Hubble profile could then be used to fit the data as the information in the core would be available.

In order to compare the effectiveness of these two methods of determining radial profiles there are two important factors to examine: the methods' precision, and their accuracy. For the precision we compare the uncertainties of the fits to the data, while for the accuracy, we compare the total numbers of objects integrated over all radii relative to the input number.

The uncertainties on the fits to the profiles of all systems with both methods were determined via bootstrapping. The radial density data for both methods was bootstrapped with 5000 iterations for each system, and powerlaw profile fitting values were determined, as well as their standard deviations. The uncertainty for each of these fitting values, the standard error, defined here in Equation \ref{eq:SE}, was determined from the standard deviations.

\begin{equation} \label{eq:SE}
    SE = \frac{\sigma}{\sqrt{n}}
\end{equation}
Here $\sigma$ is the standard deviation of the scatter of points around the fitted curve, and $n$ is the length of the dataset used to determine the fitting parameters.

\begin{figure*}[h!tb]
    \begin{center}
    \includegraphics[width=0.85\textwidth]{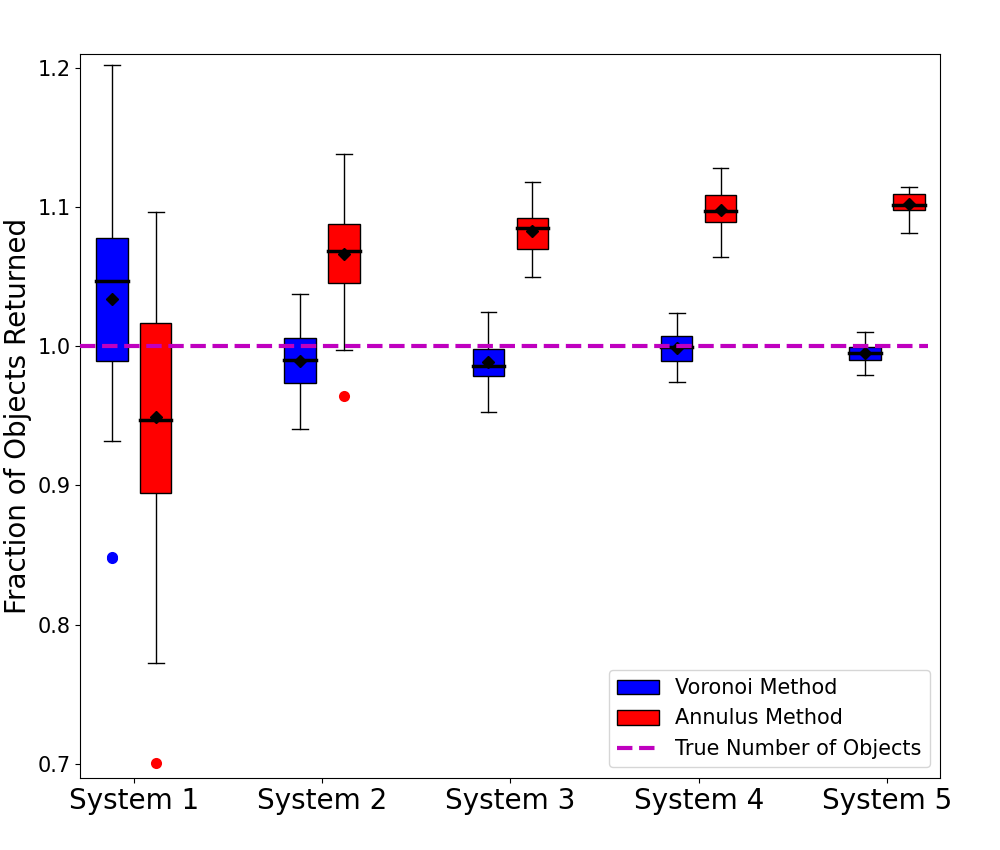}
    \caption{\label{fig:box} A boxplot depicting the results of the number of objects in each system returned when the radial density profile obtained from each method is integrated, done across all 30 random object distributions for each system. The dashed line shows where 100\% of all objects are returned. The boxes represent the interquartile range (IQR), the black lines within the boxes denote the medians, the diamonds denote the means, and the errorbars encompass all non-outlier data that lie outside the IQR. The coloured circles represent outliers, defined as data that lie greater than 1.5 times the IQR past the upper and lower quartiles}.
    \end{center}
\end{figure*}

In an ideal case, even when the total number of objects in the system is not completely observed, as in the case of the inner regions of these simulated systems seen in Figure \ref{fig:dists}, if the fit is accurate and integrated from zero to the system's maximum radius it should return the true number of objects. This was done for each run of each simulated system, and the results are displayed in Figure \ref{fig:box} as the total number of objects returned by the fit, relative to the total input number over the same radial region.

\section{Discussion} \label{sec:discussion}

It can be seen from Figure \ref{fig:box} that the Voronoi method is more successful at returning a total population that is close to the input total.  For systems 2 through 5 the Voronoi method's upper quartiles enclose or very nearly enclose the true number of objects. For these same systems however, the lower quartiles of the annulus method lie much further from the true number of objects, and for systems 3 through 5 the true number of objects is not even within their lower range. This shows that, certainly for larger systems of 2000 objects or more, the Voronoi method will more consistently and more accurately return the true number of objects.

Across Figure \ref{fig:box} we can see that the spread in both methods decreases with increasing system size. Thus, for system 1, the smallest system in this work, the spread for both methods is significantly higher than for the other systems. But even for the smallest system the Voronoi method  still returns a slightly more accurate result, with both its mean and median lying closer to the true number of objects returned than for the annulus method (0.949 and 0.947 compared to 1.033 and 1.047, respectively).

The radial profile fits for each system using each method are listed in Table \ref{tab:stats}. The precision of each of these methods is compared via the uncertainties on the fitting parameters for each method. As can be seen in Table \ref{tab:stats}, the Voronoi method consistently returns lower uncertainties on both powerlaw fitting parameters for all systems studied. As system size increases so too does the difference in method precision, with the parameter uncertainties for system 1 being comparable between methods, to systems 4 and 5 having uncertainties differing by an order of magnitude.


\begin{figure*}
    \begin{center}
    \includegraphics[width=0.95\textwidth]{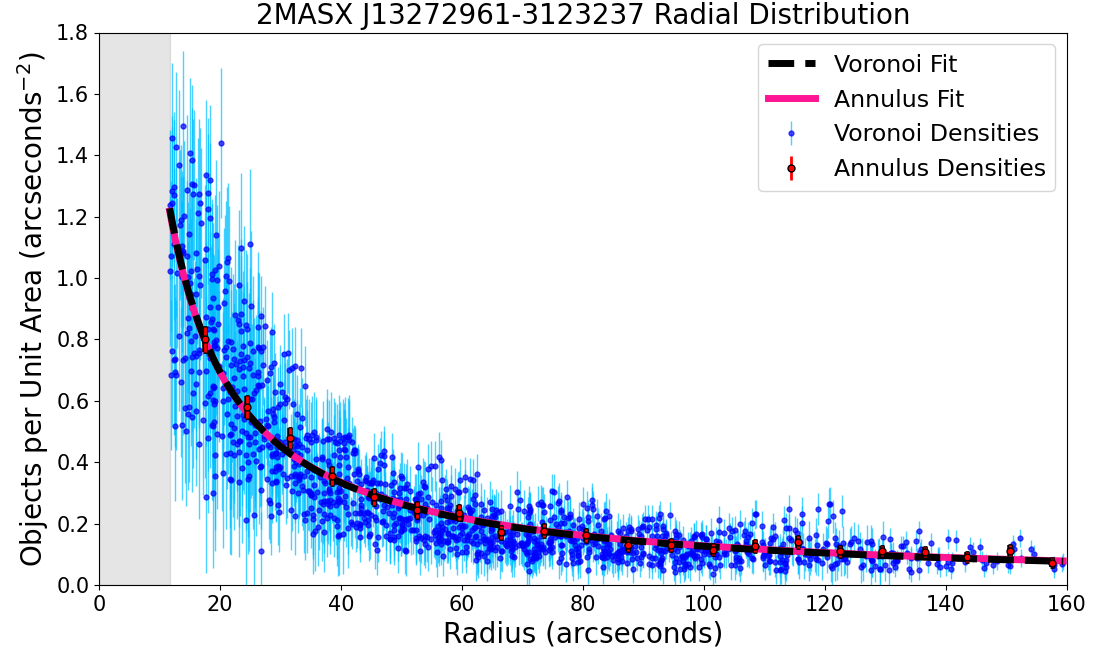}
    \includegraphics[width=0.95\textwidth]{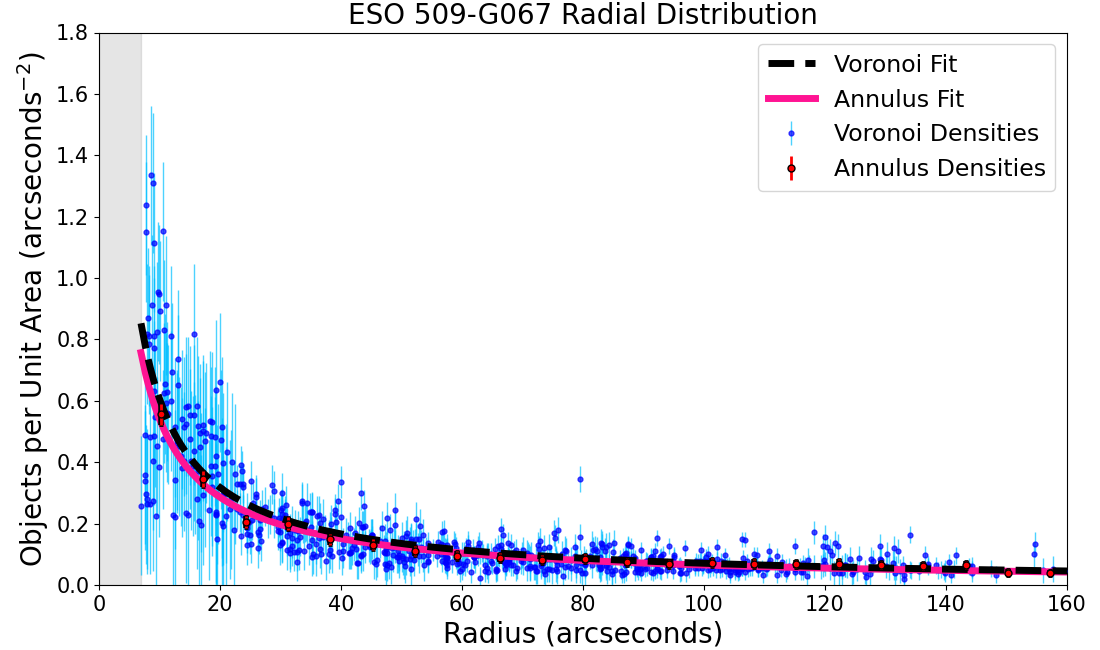}
    \caption{\label{fig:real} The density data for the observed GCSs MASX J13272961-3123237 (top) and ESO 509-G067 (bottom) using the annulus fitting method (red datapoints) and the Voronoi fitting method (blue datapoints). Both datasets are fit with a powerlaw function. The gray shaded region is unobservable due to high surface brightness.}
    \end{center}
\end{figure*}

\subsection{An Application to Observed Distributions}\label{subsec:real}

Here we present a comparison of these two methods when applied to real, observed GCSs, for which a definition of the ``true" underlying radial density profile is unknown. A common problem in many observed cases is that giant galaxies may have nearby satellites with their own GCSs, which can overlap the GCS of the central giant \citep{Dornan23, Ennis24, Lim2024}. We will return to this separate problem in followup work.  
For the purposes of this paper we will focus on two observed examples with minimal satellite interference.
The GCSs chosen for this comparison are those hosted by the BCGs 2MASSX J13272961-3123237 and ESO 509-G067. Their GCS radial density profiles have already been determined with the annulus method in Paper I, and have few significant satellite galaxies. The comparison of these two methods, applied in the identical way as for the simulated systems, can be seen in Figure \ref{fig:real}. From the work done in Paper I 2MASSX J13272961-3123237 is estimated to have a GCS population of 9100, and ESO 509-G067 to have a GCS population 4200, putting them in terms of size close to systems 3 and 4 simulated in this paper. 

The radial density fits to these observed GCSs, and the uncertainties on these fits, using both methods are shown in table \ref{tab:real_compare}. The number of GCs in the GCS and the total mass of the GCS obtained from using these fits are shown in table \ref{tab:real_compare} as well, using the same methods and photometric completion corrections as Paper I.

While no statement can be said about the comparative accuracies of these two methods when applied to observed systems, it can be seen in Figure \ref{fig:real} that the fits agree well. This is as expected as for ESO 509-G067, as for a system of this size it was found in Table \ref{tab:stats} that the fitting parameters between both methods agree within uncertainties. However, the fits between methods for 2MASSX J13272961-3123237 agree strikingly well, and is not something that is expected from the simulations for a system of its size. The further application of these methods to extremely massive observed systems is needed, and plans for this is discussed in Section \ref{sec:future}. 


The estimates of the populations themselves are roughly consistent with what we would expect based on the results from Figure \ref{fig:box}, within their uncertainties. For the larger system, 2MASSX J13272961-3123237, the annulus method returned a higher estimate than the Voronoi method, albiet with much less a dramatic difference than expected, as was seen for system 4. For ESO 509-G067, its size would make it roughly comparable to systems 2 and 3, and although the Voronoi method actually returned a higher estimate than the annulus method, within the uncertainties of the two methods, this is still be consistent with the results from Figure \ref{fig:box}. 

Overall, for both of the returned GCS values for both galaxies in Table \ref{tab:real_compare} both methods return estimates within their mutual uncertainties. However, the Voronoi method still returns estimates with lower uncertainties, indicating that this method is capable of more precise estimates.

\pagebreak

\section{Conclusions}

This work develops a new methodology to determine radial density profiles of a variety of astronomical systems of discrete objects using Voronoi tessellations. Our main conclusions are as follows:

\begin{itemize}
    \item This Voronoi method was tested and compared to the classic annulus method for five simulated systems of varying sizes and steepness of known Hubble profiles, each run 30 times with different random seeds. These systems were chosen for their similarities to observed GCSs.
    \item It was found that for all five systems the standard errors on the fits found with the Voronoi method were lower than for the fits found with the annulus method, indicating that the Voronoi method returns more precise fits regardless of system.
    \item It was found that 
    the Voronoi method returned more accurate estimates of the total population of all five systems than the annulus method. The spread in potential population estimates decreased for both methods with increasing system size.  For systems of roughly more than 1000 objects the annulus method overestimated the total population by approximately 10\%, while the Voronoi method consistently returned estimates at or within 3\% of the true value.
    \item When these methods were applied to two observed GCSs it was found that the radial density profile fits produced by the two methods were very similar and the total $N_{GC}$ estimates were within both methods' uncertainties. However, the uncertainties on both the fits and the $N_{GC}$ estimates for both observed systems were lower for the Voronoi method than for the annulus method.
\end{itemize}

\begin{center}
\begin{table*}
    \caption{Comparison of Annulus and Voronoi Methods on Observed Galaxies} \label{tab:real_compare}
    \begin{tabular}{|c|c|c|c|}
    \hline \hline
    Galaxy & Value & Annulus Method & Voronoi Method \\
    \hline
    \multirow{2}{*}{J13272961} & Powerlaw & $\sigma =  (16.1 \pm 1.5) r^{(-0.92 \pm 0.03)}$ & $\sigma =  (16.6 \pm 0.8) r^{(-0.93 \pm 0.01)}$\\
     & $N_{GC}$ & $9055 \pm 1173$ & $8885 \pm 793$\\
    \hline \hline \hline
    \multirow{2}{*}{ESO 509-G067} & Powerlaw & $\sigma =  (4.5 \pm 0.5) r^{(-0.92 \pm 0.03)}$ & $\sigma =  (5.2 \pm 0.02) r^{(-0.93 \pm 0.01)}$\\
     & $N_{GC}$ & $4174 \pm 726$ & $4632 \pm 471$\\
    \hline
    \end{tabular}
\end{table*}
\end{center}

\subsection{Future Work}\label{sec:future}

Future work with this new method of radial density profile fitting will include its application to more observed systems of varying sizes. Currently the authors of this paper plan to apply this method to the remaining GCSs studied in Paper I, as well as 17 additional BCGs imaged with the HST. 

Further improvements to this method also include the ability to accurately remove objects associated with nearby systems. As briefly mentioned in section \ref{subsec:real}, observed GCSs often have nearby satellite galaxies with their own GCSs which can contaminate radial density profile estimates of the target system. The current method of removal of these contaminating systems is to simply mask out any objects suspected to not belong to the target system \citep{Dornan23, Lim2024}, or even to simply leave the contaminating system if it is small enough to not have a drastic affect on the target system \citep{Harris15, Mulholland17}. This method can lead to objects associated with the target system being removed, or missing various objects associated with the contaminating system.

With the increase in spatial density data associated with the Voronoi method compared to the annulus method, it is possible that the radial density profile of the contaminating system, and its influence on the estimated profile of the target system, can be determined and subtracted from the target system's fit. This would not only give a far more accurate estimate of the profile for systems in dense, clustered environments, but would also allow for an estimate of the smaller contaminating system's radial density profile. This can be particularly helpful in situations where the GCS distributions of two interacting galaxies is trying to be studied \citep[eg.][]{Ennis24}. Work on developing this method is in progress.

\section{Acknowledgements}

This work was supported by a Discovery Grant to WEH from the Natural Sciences and Engineering Research Council of Canada (NSERC).

\bibliography{paper}{}
\bibliographystyle{aasjournal}

\end{document}